%% file: paper.tex
\input{macros.tex}

\def\j{\mbf j}
\def\etal{{\it et al. }}
\documentclass[twocolumn,showpacs,preprintnumbers,amsmath,amssymb,prb]{revtex4}

\usepackage{graphicx}
\usepackage{dcolumn}
\usepackage{bm}
\begin{document}

\title{Optical properties of random alloys  : Application to Cu$_{50}$Au$_{50}$ and Ni$_{50}$Pt$_{50}$}

\author{Kamal Krishna Saha}
\altaffiliation{kamal@bose.res.in}
\author{Abhijit Mookerjee}
\altaffiliation{abhijit@bose.res.in}
\affiliation{ S. N. Bose National Centre for Basic Sciences. Block-JD, Sector-III, \\
Salt Lake City, Kolkata-700098, India.}

\begin{abstract}
{In an earlier paper [K. K. Saha and A. Mookerjee, \prb {\bf 70} (2004) (in press) or, cond-mat/0403456] 
we had presented a  formulation for the calculation of the 
configuration-averaged optical conductivity in 
random alloys. Our formulation is based on the augmented-space theorem introduced by one of us 
[A. Mookerjee, J. Phys. C: Solid State Phys. {\bf 6}, 1340 (1973)].  In this communication we shall
combine our formulation with the tight-binding linear muffin-tin orbitals (TB-LMTO) 
technique to study the optical conductivities of two alloys Cu$_{50}$Au$_{50}$ and Ni$_{50}$Pt$_{50}$.}
\end{abstract}
\date{\today}

\pacs{71.23.-k}
\maketitle
\parindent 0pt

\section{Introduction}
In this communication we propose to study the optical properties of disordered CuAu and NiPt 50-50 alloys from a first principle approach. We have chosen these two alloy systems because of several reasons : for CuAu, the bunch of $d$-like states sits about 1 eV below the Fermi level. For low photon energies, therefore, optical conductivity is dominated by the
intra-band transitions within $s-p$-like states, which are extended and rather free electron-like. As a consequence, the optical conductivity for low photon energies below $\simeq $ 1 eV should have a Drude like behaviour. For higher photon 
 energies inter-band transitions between the occupied $d$-states and the higher unoccupied states begin to take over. 
In sharp contrast, the Fermi energy of NiPt almost straddles the $d$-like peak. For this alloy the Drude behaviour 
should be confined to a
very narrow low photon energy range. This contrasting behaviour should be reflected in our results. Moreover, in both
the two alloy systems there is a large size mismatch between the constituents. This indicates that the standard
single site mean-field theories would be inadequate to capture the effect of this large size-mismatch. These alloy systems are therefore ideal to illustrate the advantages of the augmented-space recursion (ASR) method proposed by us \cite{Am2}.

Earlier theoretical work on optical conductivity for random alloys began with Velick\'y \cite{vel} based on the single site coherent potential approximation (CPA) in an empirical tight-binding model alloy. Butler \cite{but} extended
the ideas and combined the CPA with the first-principles Korringa-Kohn-Rostocker (KKR) technique. Banhart \cite{ban}
used the KKR-CPA to study the optical conductivity of AgAu alloys. This alloy system has close resemblance to CuAu.
Banhart found discrepancies of his theoretical results with experiment \cite{riv,niel} and argued that various factors could be responsible : use of the density functional and  the single-site mean-field approximations in theory and 
effects of surfaces, their roughness, possible adsorbates and presence of large stresses in the samples, in the
experiments. There have been a few more theoretical studies of optical properties of random alloys :  Rhee \etal \cite{rhee} on CoAl, Uba \etal \cite{uba} on CoPt and  Rhee \etal \cite{rhee2} on Ni$_3$Al.  These works all base their
approach on a large super-cell method to take care of the disorder. The method is brute force and less satisfactory than the CPA or ASR. 

In an earlier paper \cite{opt} we had presented a formulation for obtaining the configuration averaged optical
conductivity for random binary alloys. 
We had chosen as our basis the minimal set of the tight-binding linear muffin-tin orbitals method \cite{Ander1,Ander2}.
Configuration averaging over various random atomic arrangements had been carried out
using the augmented-space  formalism   (ASF) introduced by us earlier for the study of electronic properties of
disordered systems \cite{Am,Am3}. The ASF goes beyond the usual single site 
mean-field approaches and takes into account configuration fluctuations about the mean-field. We shall present here a
summary of the results derived in the earlier paper \cite{opt}. 
In linear response theory, at zero temperature, the real part of the optical conductivity of a disordered alloy is 
given by the Kubo-Greenwood expression~:
\be
\sigma(\omega) \ =\ \frac{S(\omega)}{\omega}
\ee

\n where, the configuration averaged current-current correlation function $\ll S(\omega)\gg$ is given by :
\begin{eqnarray}
\frac{1}{3\pi}\sum_{\gamma}\mathrm{Tr}\int dE\left\langle\rule{0mm}{4mm}{\mathbf j}_{\gamma}
\mbox{Im} \{ {\mathbf G}^v ( E)\} {\mathbf j}_{\gamma}^\dagger\ 
\mbox{Im} \{{\mathbf G}^c (E+\omega)\} \right\rangle 
\label{ref1}
\end{eqnarray}
\noindent If we define
\be 
\ll{S}_\gamma(z_1,z_2)\gg  = \mbox{Tr}\ \left\langle\ \rule{0mm}{4mm}{\mathbf j}_{\gamma}\ {\mathbf G}^v (z_1)\ 
{\mathbf j}_{\gamma}^\dagger\ {\mathbf G}^c (z_2)\right\rangle
\label{eq4}
\ee

\noindent  then, using the Herglotz properties of the Green function, the correlation function becomes
\begin{eqnarray}
\ll S(\omega)\gg = \frac{1}{12\pi}\sum_{\gamma}\ \int dE\ \left[\ {\cal S}_\gamma(E^-,E^{+}+\omega)\right. \nonumber\\
\phantom{xx} + {\cal S}_\gamma(E^+,E^{-}+\omega)-{\cal S}_\gamma(E^+,E^{+}+\omega)\nonumber\\  
\left. -{\cal S}_\gamma(E^-,E^{-}+\omega)\right]\phantom{XXXXX}
\end{eqnarray}

\noindent where
\[ f(E^\pm) \eq \lim_{\delta\rightarrow 0} f(E\pm i\delta). \]

\n The formulation we presented in our earlier paper \cite{opt} was based on the disordered induced
scattering leading to the renormalization of the propagator ${\mathbf G}$ as well as the current
terms ${\mathbf j}$. The dominant contributions to the configuration averaged correlation function 
was due to joint fluctuations of an averaged current and two electron propagators :
\begin{widetext}
\begin{eqnarray}
\ll {S}^{(0)}_\gamma(z_1,z_2)\gg \ = 
\sumk \ll {\mathbf \j}_\gamma(\k)\gg\ \ll {\mathbf G}^v(\k,z_1)\gg\ \ll {\mathbf \j}_\gamma(\k)\gg^\dagger\ \ll {\mathbf G}^c(\k,z_2)\gg.
\end{eqnarray}
\end{widetext}
\n We have also shown that disorder scattering renormalizes that averaged current to an effective current
term and the average propagator to a configuration averaged propagator beyond the CPA approximation. We
expressed the effective current in terms of the self-energy as :
\begin{widetext}
\begin{eqnarray}
\mbf{J}^{\rm{eff}}_\gamma(\k,z_1,z_2)& = \ \ll {\bf j}_\gamma(\k)\gg + 2\left[\rule{0mm}{5mm} \mbf{\Sigma}(\k,z_2)\
\mbf{f}(z_2)\ {\j^{(1)}_\gamma}^\dagger(\k) \pls {\j^{(1)}_\gamma}^\dagger(\k)\ \mbf{f}(z_1)\ \Sigma(\k,z_1)\right] \nonumber \\
&  \pls \Sigma(\k,z_2)\ \mbf{f}(z_2)\ \j^{(2)}_\gamma(\k)\ \mbf{f}(z_1)\ \Sigma(\k,z_1).
\label{eq6}
\end{eqnarray}

\n and the contribution to the effective current to the correlation function was :
\begin{eqnarray}  \ll {S}^{(1)}_\gamma(z_1,z_2)\gg\ \eq\ \sumk\ \mbox{Tr}\left[\rule{0mm}{4mm}
\mbf{J}^{\rm{eff}}_\gamma(\k, z_1,z_2)\ll \mbf{G}^v(\k,z_1)\gg
\mbf{J}^{\rm{eff}}_\gamma(\k,z_1,z_2)^{\dagger}\ll \mbf{G}^c(\k,z_2)\gg\right].
\end{eqnarray}
\n The contribution of joint fluctuations between the two current terms and one propagator
was given by
\begin{eqnarray}
\ll {S}^{(2)}_\gamma(z_1,z_2)\gg  &\eq& 4\ \sumk \ \mbox{Tr} \left [ \ \rule{0mm}{5mm} \j^{(1)}_\gamma(\k)\ \mbf{f}(z_1)\ \Sigma(\k,z_1)\ \mbf{f}(z_1)\ {\j^{(1)}_\gamma}^\dagger(\k) \ \ll \mbf{G}(\k,z_2)\gg \right.\nonumber\\
&& \qquad \left. \pls {\j^{(1)}_\gamma}^\dagger(\k) \ \mbf{f}(z_2)\ \Sigma(\k,z_2)\ \mbf{f}(z_2)\ \j^{(1)}_\gamma(\k)\ \ll\mbf{G}(\k,z_1)\gg \rule{0mm}{5mm}\right].
\end{eqnarray}
The vertex correction terms in the ladder approximation contributed :
\begin{eqnarray}
\ll S^{\rm{ladder}}_\gamma(z_1,z_2)\gg \ =\ \mbox{Tr}
\sum_{L_1L_2}\sum_{L_3L_4}\Gamma_\gamma^{L_1L_2}(z_1,z_2)\ \Lambda^{L_1L_3}_{L_2L_4}\ 
\widehat{\Gamma}^{L_3L_4}_\gamma(z_1,z_2)
=\ \mbox{Tr} \ \mbf{\Gamma}_\gamma(z_1,z_2)\otimes \mbf{\widehat{\Gamma}}_\gamma(z_1,z_2)\ {\Lambda}(z_1,z_2). \quad\qquad
\label{eq9}
\end{eqnarray}
\end{widetext}
\n where
\begin{eqnarray*}
\int_{\rm{BZ}} \frac{d^3\k}{8\pi^3}\ \mbf{G}(\k,z_2)\ \mbf{J}^{\rm{eff}}_\gamma(\k,z_1,z_2)\ \mbf{G}
(\k,z_1) = \mbf{\Gamma}_\gamma(z_1,z_2) \qquad\\
\int_{\rm{BZ}} \frac{d^3\k'}{8\pi^3}\ \mbf{G}(\k',z_1)\
\mbf{J}^{\rm{eff}}_\gamma(\k',z_1,z_2)\!\!\phantom{i}^{\dagger} \mbf{G}(\k',z_2)=\widehat{\mbf{\Gamma}}_\gamma(z_1,z_2). \quad
\end{eqnarray*}

\n and
\begin{eqnarray*}
\lambda^{L_1L_2}_{L_3L_4}(z_1,z_2) = \int_{\rm{BZ}}\ \frac{d^3\k}{8\pi^3}\ 
G_{L_3L_4}(\k,z_1)\ G_{L_2L_1}(\k,z_2)\phantom{XXXX} \\
\omega^{L_1L_2}_{L_3L_4} =  W^{L_1}_{L_3}\ \delta_{L_1L_2}\ \delta_{L_3L_4}\phantom{XXXXXXXXXXXXXXXX}
\end{eqnarray*}

\begin{eqnarray*}
W^L_{L'}  = F_L(z_2) \left[\rule{0mm}{3mm} \delta_{LL'}+2 \sum_{L^{''}} \left[\rule{0mm}{3mm} B_{L^{''}}(z_1) \ G_{RL^{''},RL'}(z_1) \right.\right.\phantom{XX} \\
\left.\left. \ +B_{L^{''}}(z_2)\  G_{RL^{''},RL'}(z_2)\right]\right]F_{L'}(z_1). \qquad \\
F_L(z) = \sqrt{xy}\ d\left(\frac{C_L-z)}{\Delta_L}\right)/\ll 1/\Delta_L\gg \phantom{XXXXXXXX} \\
B_L(z) = (y-x)\ d\left(\frac{C_L-z)}{\Delta_L}\right)/\ll 1/\Delta_L\gg\phantom{XXXXXXX}
\end{eqnarray*}

\n Here $x,y$ are the concentrations of the component atoms, $C$,$\Delta$ are the standard potential parameters of 
the TB-LMTO and $d(f) = f_A -f_B$. These super-matrices in $\{L\}$ space are written as $\uu\lambda$ and $\uu\omega$. 
The full ladder vertex may now be written as
\begin{eqnarray}
\uu\Lambda(z_1,z_2) &=&\uu\omega+ \uu\omega\ \uu\lambda\ \uu\omega + \uu\omega\ \uu\lambda
\ \uu\omega\ \uu\lambda\ \uu\omega + \ldots\nonumber\\
 &=& \uu\omega \left( \uu I \mns \uu\lambda(z_1,z_2)\ \uu\omega \right)^{-1}
\end{eqnarray}

Including all sorts of corrections the averaged correlation function is then :

\begin{eqnarray*}
\ll{S}_\gamma(z_1,z_2)\gg = \ll{S}_\gamma^{(1)}(z_1,z_2)\gg+\ll{S}_\gamma^{(2)}(z_1,z_2)\gg  \\
 + \ll{S}_\gamma^{\mathrm{ladder}}(z_1,z_2)\gg\phantom{XXXX}
\end{eqnarray*}

\section{Results and Discussion}

We have begun our study with the self-consistent TB-LMTO-ASR calculation on NiPt and CuAu 50-50 alloys. We have
minimized the energy with respect to the variation in the average lattice constant for both the alloys. The table
shows the lowest energy lattice constants and compares them with the averaged or Vegard's law results. As expected,
because of the large size difference between the constituents there is a ``bowing" effect which is most prominent
at the 50-50 alloys. The lowest energy lattice constant for both the alloys is greater than the Vegard's law
predictions.

\begin{table}
\centering
\begin{tabular}{ccc} \hline\hline
 & Lowest energy & Vegard's Law \\
Alloy      & lattice const & lattice const \\
      & (\r{A})          & (\r{A})       \\ \hline
Cu$_{50}$Au$_{50}$ & 7.31 & 7.26 \\
Ni$_{50}$Pt$_{50}$ & 7.09 & 7.03 \\
\hline
\end{tabular}
\caption{Lowest energy and Vegard's Law lattice constants for CuAu and NiPt.}
\end{table}

\begin{figure}
\includegraphics[width=3.0in, height=2in]{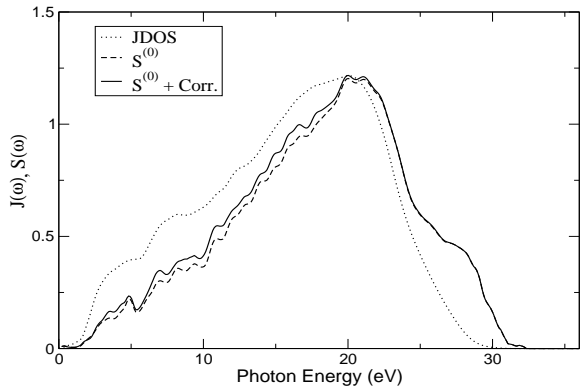}
\caption{\label{fig1} 
The configuration averaged joint density of states and correlation function
for CuAu (50-50) alloy shown as a function of the photon energy.}
\end{figure}

Figure \ref{fig1} shows the comparison between the scaled joint density of states (JDOS) and the averaged correlation
function for a CuAu (50-50) alloy. From the figure it is clear that the transition rate is dependent
both on the initial and the final energies, throughout the frequency range of interest. That is :
\[S(\omega) \neq \vert T\vert^2 \ J(\omega)\]
\n where 
\[J(\omega) = \int dE\int\ \frac{d^3\k}{8\pi^3}\ \mathrm{Tr} \langle \mathbf{G}^c(\k,E)\mathbf{G}^v(\k,E+\omega)\rangle. \]
The Fig.~\ref{fig1} also shows that the disorder corrections to the current and the vertex correction
are rather small and become negligible beyond photon energies of about 22 eV.
\begin{figure}
\includegraphics[width=3.0in, height=2in]{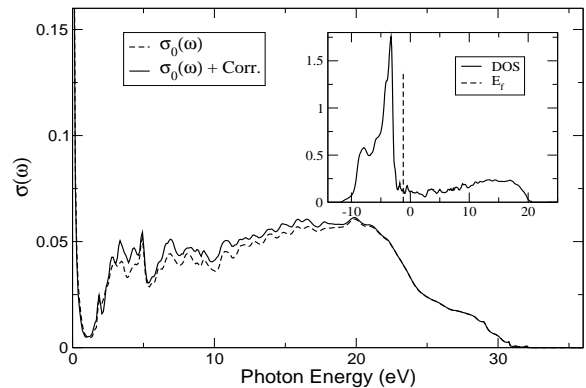}
\caption{\label{fig2}
Averaged optical conductivity, $\sigma_0(\omega) = S^{(0)}/\omega$, and the density of states for a CuAu (50-50) alloy.
The corrections to the optical conductivity are shown in Eqs.~(\ref{eq6}-\ref{eq9}).} 
\end{figure}

Figure \ref{fig2} shows the optical conductivity $\sigma(\omega)$ for CuAu (50-50) alloy. 
The inset shows the configuration 
averaged density of states for the same alloy. The edge of the $d$-band complex is clearly seen to lie about
1 eV below the Fermi energy. The optical conductivity rapidly decreases as we increase the photon energy
from zero upwards. This decrease continues until about 1 eV and then the conductivity rises again and has
considerable structure as also shown in the correlation function for these photon energies (Fig.~\ref{fig1}).
\begin{figure}
\includegraphics[width=3.0in, height=2in]{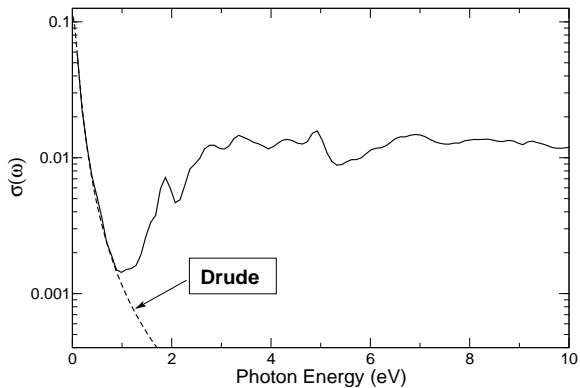}
\caption{\label{fig3}
Averaged optical conductivity showing a Drude fit at low photon energies.}
\end{figure}

Figure \ref{fig3} shows the optical conductivity with a Drude fit [$\sigma^D(\omega)=\sigma(0)/(1+{(\omega\tau)}^2),
\ \mathrm{with} \ \sigma(0)=0.11,\ \tau=9.78 $] for the lower photon energies. The
Drude fit is good for photon energies below 1 eV. From this information we may deduce that for low photon energies the conductivity arises due to intra-band transition between the $s-p$ states, which are free electron like and lead to a Drude type behaviour. Above 1 eV there is a onset of inter-band transition between the $d$ and the conduction states and this leads to a sharp increase of optical conductivity and structure reflecting
the structures in the $d$-like states.
\begin{figure}
\includegraphics[width=3.0in, height=2in]{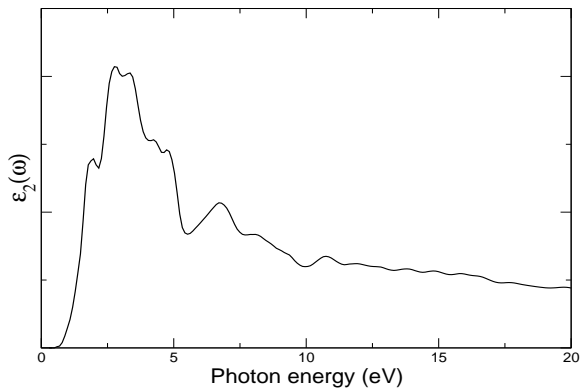}
\caption{\label{fig4}
Inter-band contribution to the imaginary part of the dielectric function for CuAu (50-50) alloy.}
\end{figure}
The inter-band contribution to the imaginary part of the dielectric function $\epsilon_2(\omega)$ may be
obtained from the optical conductivity data, by subtracting away the Drude contribution and dividing the 
result by $\omega$ : $\epsilon_2(\omega) = (\sigma(\omega)-\sigma^D(\omega))/\omega$ . Below the onset of the
inter-band transitions, this quantity vanishes. It then reaches a maximum at around 3 eV before decreasing.
We have experimental data on AgAu (50-50) \cite{niel}, whose density of states closely resembles CuAu.
The experimental data are in good qualitative agreement with  Fig.~\ref{fig4}.
The general shape with a shoulder around 1eV, a maximum and around 3 eV is clearly reproduced.
\begin{figure}
\includegraphics[width=3.0in, height=2in]{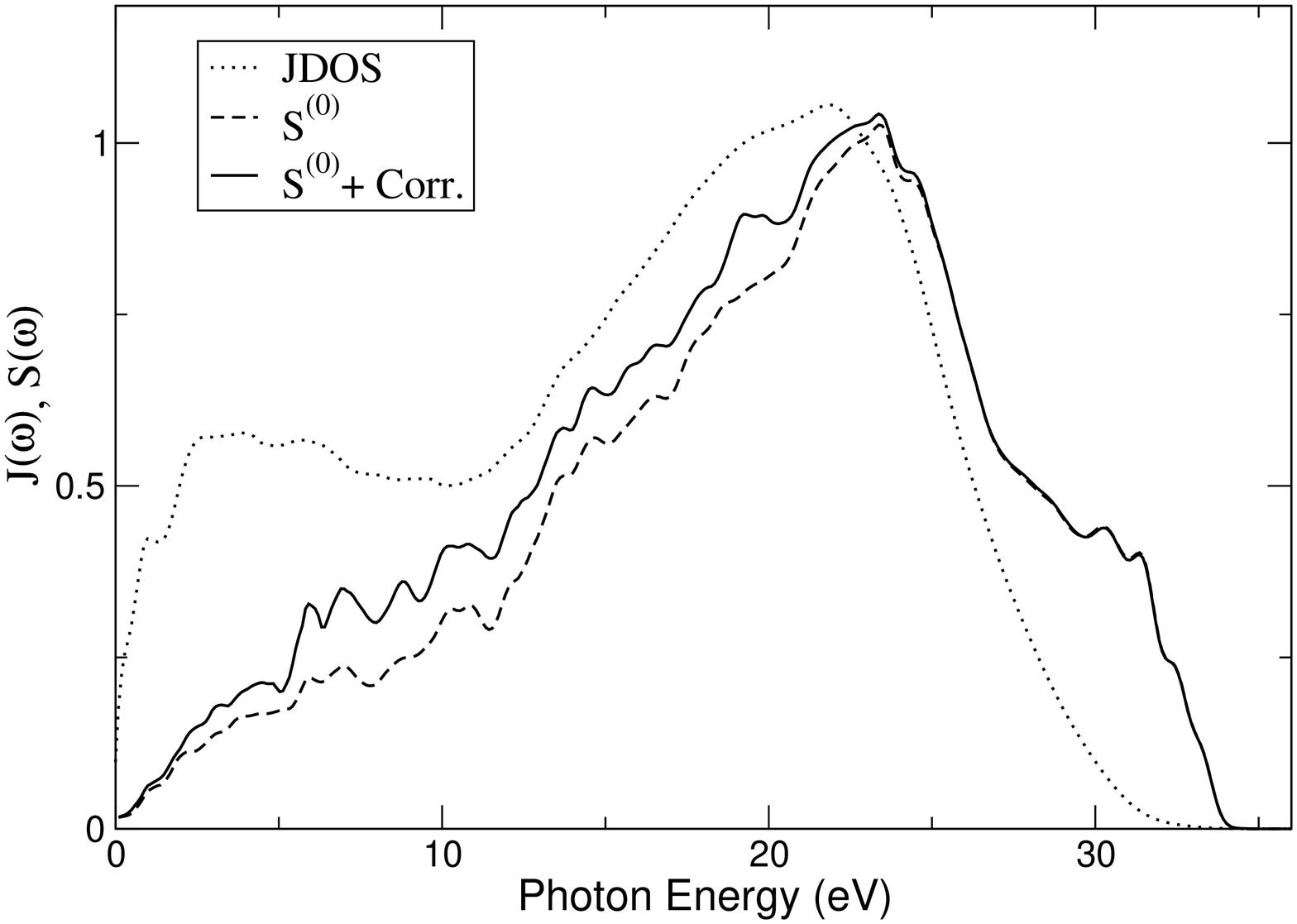}
\caption{\label{fig5}
The configuration averaged joint density of states and correlation function for NiPt (50-50) alloy shown as a function of photon energy.}
\end{figure}

Figure~\ref{fig5} shows the joint density of states and the averaged correlation function for the NiPt (50-50) alloy. The energy-frequency dependence of the effective transition rate is considerable more pronounced
than for CuAu. Disorder correction to the current terms and vertex corrections are also more in the low photon energy region. They become negligible for  high photon energies.
\begin{figure}
\includegraphics[width=3.0in, height=2in]{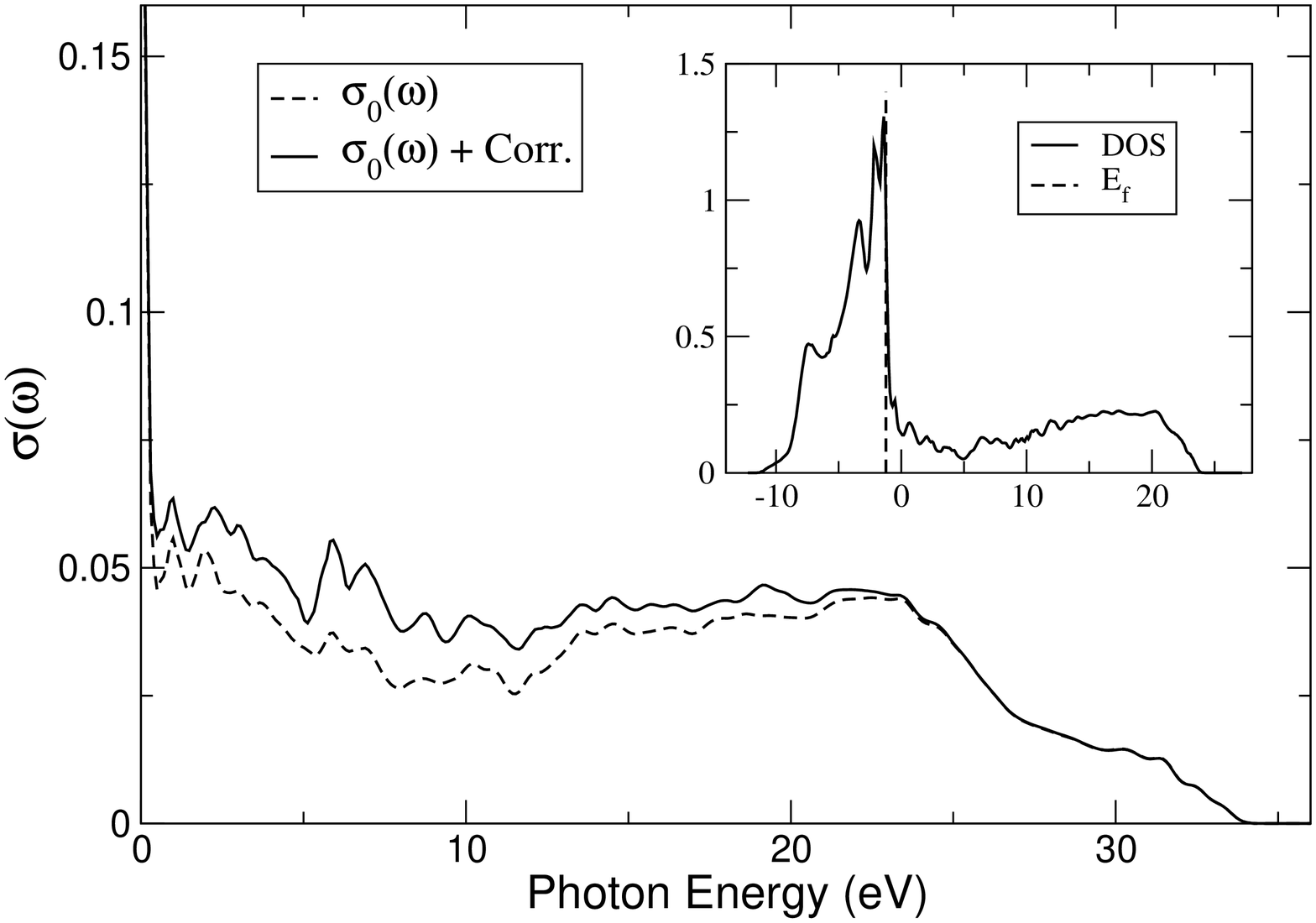}
\caption{\label{fig6}
Averaged optical conductivity and the density of states for NiPt (50-50) alloy.}
\end{figure}

Figure \ref{fig6} shows the density of states and the averaged optical conductivity for NiPt. Although the density of states for NiPt qualitatively resembles that for CuAu, unlike the latter, the Fermi level sits right atop the high peak due to the $d$-like states. The inter-band transitions between the $d$-states and the conduction band is expected to start for very small photon energies, with a Drude contribution confined to a very narrow energy range near zero. The optical
conductivity falls sharply in a very narrow energy range and recovers almost immediately. This is expected
from the density of states picture. Since the Drude fit is in a very narrow range indeed we do not show it explicitly in the figure.
\begin{figure}
\includegraphics[width=3.0in, height=2in]{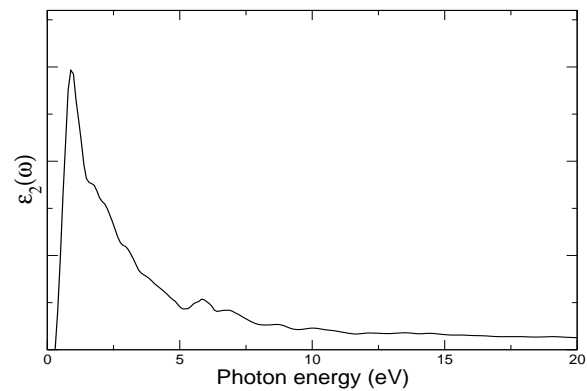}
\caption{\label{fig7}
Inter-band contribution to the imaginary part of the dielectric function for NiPt (50-50) alloy.}
\end{figure}

In Fig.~\ref{fig7} we show the inter-band contribution to imaginary part of the dielectric function
for NiPt (50-50) alloy. The inter-band contribution begins at a very low photon-energy as expected and
attains a maximum around 1 eV. This is in contrast to the behaviour of CuAu, where Drude behaviour 
persists over a longer energy interval. We were unable to locate experimental data for this alloy system 
for comparison.

\section{Conclusion}

The frequency dependent transport quantities of two disordered metallic alloys have been investigated 
applying a TB-LMTO-ASR based first principle theory \cite{opt}. We have found that the energy-frequency 
dependence to the effective transition rate is more pronounced for NiPt than that of CuAu. Same tendency
was observed in disordered correction to the current terms and vertex corrections in the low photon 
energy region. We have also found that the conductivity occurs because of both intra-band and inter-band 
transitions. So the imaginary part of the dielectric function calculated from the conductivity splits 
into intra and inter-band contributions. For CuAu the intra-band transition takes place up to $\simeq$~1 eV 
photon energy and in this region the conductivity curve follows Drude law of free electron model. 
For NiPt alloy the Drude behaviour is confined to a very narrow energy range, as the Fermi energy straddles 
the $d$-like peak in density of states. We have compared our results with available theoretical 
and experimental results and achieved a very impressive agreement with them.

\end{document}

%% file: macros.tex
\def\sumk{\int_{\rm{BZ}}\ \frac{d^3\k}{8\pi^3}\ }
\def\uu#1{\underline{\underline{#1}}}

\def\ket{\vert \vert  \{ \emptyset \} \rangle}
\def\ket2{\vert \vert \otimes \{ R \} \rangle}

\def\n{\noindent }
\def\mbf#1{{\mathbf {#1}}}

\def\eq{\ =\ }
\def\mns{\ -\ }
\def\pls{\ +\ }
\def\be{\begin{equation}}
\def\ee{\end{equation}}
\def\<{[}
\def\>{]}

\def\k{{\mathbf k}}